\documentclass[preprint,showpacs,preprintnumbers,amsmath,amssymb,nofootinbib]{revtex4}

\usepackage{graphicx}% Include figure files
\usepackage{dcolumn}% Align table columns on decimal point
\usepackage{bm}% bold math

\newcommand{\bear}{\begin{eqnarray}}    
\newcommand{\eear}{\end{eqnarray}}      
\newcommand{\beqstar}{\begin{eqnarray*}}        
\newcommand{\eeqstar}{\end{eqnarray*}}

\def\eslt{E_T^{miss}}
\def\to{\rightarrow}

\def\bi{\begin{itemize}}
\def\ei{\end{itemize}}

\def\tu{\tilde u}
\def\sps1ap{SPS1a$^\prime$}
\def\c1p{C1$^\prime$}

\def\ttau{\tilde \tau}

\def\tg{\tilde g}

\def\tq{\tilde q}
\def\tw{\widetilde W}
\def\tz{\widetilde Z}
\def\alt{\stackrel{<}{\sim}}
\def\agt{\stackrel{>}{\sim}}
\def\be{\begin{equation}}  
\def\ee{\end{equation}}  
\def\bea{\begin{eqnarray}}  
\def\eea{\end{eqnarray}}  
\def\beas{\begin{eqnarray*}}  
\def\eeas{\end{eqnarray*}}  
\newcommand\plb[3]{{Phys.\ Lett.\ }{\bf B #1}, #2 (#3)}
\newcommand\jhep[3]{{J. High Energy Phys.\ }{\bf #1}, #2 (#3)}

\newcommand\zpc[3]{{Z.\ Physik }{\bf C #1}, #2 (#3)}
\newcommand{\hepph}[1]{hep-ph/#1}

\begin{document}

%\preprint{EFI-02-80, UFIFT-HEP-02-7, CERN-TH/2002-108, BUHEP-02-21}

\title{Early SUSY discovery at LHC via sparticle \\
cascade decays to same-sign and multimuon states}

\author{Howard Baer}
\affiliation{Dept. of Physics and Astronomy,
University of Oklahoma, Norman, OK 73019, USA}
\author{Andre Lessa}
\affiliation{Dept. of Physics and Astronomy,
University of Oklahoma, Norman, OK 73019, USA}
\author{Heaya Summy}
\affiliation{Dept. of Physics and Astronomy,
University of Oklahoma, Norman, OK 73019, USA}

\date{January 16, 2009}

\begin{abstract}
In the very early stages of LHC running, uncertainties in detector
performance will lead 
to large ambiguities in jet, electron and photon energy measurements,
along with inferred missing transverse energy $\eslt$. However, muon detection
should be quite straightforward, with the added benefit that muons
can be reliably detected down to transverse energies of order 5 GeV. 
Supersymmetry discovery through multimuon channels has been extensively 
explored in the literature, but always relying on hard $\eslt$ cuts. Here, we
quantify signal and background rates for same-sign (SS) dimuon and multimuon 
production at the LHC without any $\eslt$ cuts. The LHC, operating at
$\sqrt{s}=10$ TeV,  should be able to discover
a signal over expected background consistent with gluino pair production 
for $m_{\tg}\alt 450$ ($550$) GeV in the SS dimuon plus $\ge 4$ jets 
state with just 0.1 (0.2) fb$^{-1}$ of integrated luminosity.
\end{abstract}

\maketitle

With the recent circulation of proton beams around the entire 
CERN Large Hadron Collider (LHC) ring, 
the era of LHC physics has begun. 
Meaningful data is now expected starting in fall 2009, when LHC will likely
start up with $pp$ collisions at $\sqrt{s}\simeq 10$ TeV.

In the LHC ramping up process, it will be essential to observe 
many familiar Standard Model (SM) processes--
multi-jet production as predicted by QCD, $W$ and $Z$ production
as predicted by the electroweak theory, $t\bar{t}$
production, vector boson pair production--
all at their expected rates, and with distributions and mass peaks 
at previously measured values\cite{bg}.
Conventional wisdom holds that once confidence in the Atlas and CMS 
detectors has been established, 
then the search for physics {\it beyond the Standard Model} will begin. 
In this letter, we explore the possibility of searching for new physics 
{\it in parallel} with the calibration phase. We will show that even with 
relatively poor knowledge of the detector, new physics searches
may still be possible, at least in the case of weak scale supersymmetry. 

Weak scale supersymmetry-- wherein each particle state of the
Standard Model has a TeV-scale superpartner differing by $1/2 \hbar$ 
units of spin--
is perhaps the most motivated new physics theory\cite{susy}. 
Theories with supersymmetry (SUSY) broken at the weak scale 
actually enjoy {\it indirect} experimental
support in that the measured values of the three SM gauge couplings
at energy scale $Q\simeq M_Z$, 
when extrapolated to very high energies under renormalization group (RG) 
evolution, meet at a point as predicted by grand unified theories (GUTs)
under Minimal Supersymmetric Standard Model (MSSM) evolution (while they miss
badly under SM evolution)\cite{gcevol}. SUSY theories also predict a SM-like Higgs boson $h$
with mass below $\sim 135$ GeV-- a scenario which is consistent with
global analyses of precision electroweak measurements\cite{ew}.

While the idea of supersymmetry is theoretically very appealing, 
the mechanism behind SUSY breaking is a complete mystery. One very elegant
SUSY breaking mechanism occurs in {\it local} SUSY-- or supergravity (SUGRA)--
theories. It is possible to embed the SM into a supergravity theory, 
and then set up a {\it hidden sector} which serves as an arena for
SUSY breaking via the super-Higgs mechanism. The SUSY breaking is communicated 
from the hidden sector to the visible sector via Planck-scale suppressed
operators, and a judicious choice of parameters leads to weak scale
soft SUSY breaking (SSB) parameters, 
exactly as needed by gauge coupling evolution,
and which serve to stabilize the weak scale-GUT scale energy hierarchy
without too much fine-tuning. The simplest such model, the minimal
supergravity or mSUGRA model, thus posits a common (universal) 
mass $m_0$ for all SSB scalar masses, a common SSB gaugino mass
$m_{1/2}$, and common trilinear SSB terms $A_0$. 
(Here, the gaugino is the spin-$1\over 2$ superpartner of the gauge bosons.)
Motivated by gauge coupling unification, these common masses are assumed valid at the 
GUT scale $M_{GUT}\simeq 2\times 10^{16}$ GeV. All weak-scale Lagrangian parameters
can be calculated in terms of this parameter set using the power of the RG equations.
Thus, all physical superpartner masses and mixings may be calculated in terms of the
parameter set 
\be
 m_0 ,\ m_{1/2} ,\ A_0 ,\ \tan\beta , sign (\mu ) ,
\ee
wherein $\tan\beta$ is the ratio of the two Higgs field vacuum expectation values (vevs) needed for electroweak
symmetry breaking, and $\mu$ is a quadratic superpotential term.
While many other well-motivated SUSY models exist, the mSUGRA model
%\footnote{In the
%literature, the mSUGRA model is sometimes also referred to (as of 1994) as the CMSSM, 
%or constrained MSSM. 
%Since that time, numerous other constrained MSSMs have emerged, so CMSSM is no longer 
%uniquely specified.} 
has emerged as a sort of
paradigm choice for exploring basic SUSY phenomena expected at collider experiments.

The strongly interacting sparticles-- the gluinos $\tg$ and squarks $\tq$-- 
often end up with the 
largest of all the sparticle masses due to the influence of the strong interactions 
on their RG mass evolution.
Sparticles such as charginos, neutralinos and sleptons are frequently much lighter. The strongly 
interacting $\tg$ and $\tq$-- produced through QCD interactions-- 
usually have the largest production cross sections. 
Once the $\tg$ and $\tq$s are produced, they decay through a cascade of possibly several 
stages until the state with the lightest SUSY
particle-- or LSP-- is reached\cite{cascade}. The LSP in mSUGRA usually turns out to be the lightest
neutralino, $\tz_1$, which if $R$-parity is conserved, is absolutely stable and serves as a good
candidate for cold dark matter (CDM) in the universe\cite{haim}.

The classic signature for $\tg$ and $\tq$ production at hadron colliders consists of
events containing jets plus large missing transverse energy $\eslt$, wherein the 
$\eslt$ arises due to the $\tz_1$s completely escaping the detector, much as neutrinos do.
This signature channel should serve sparticle-hunters well once the detectors are fully
calibrated so that SM backgrounds for jets$+\eslt$ events are well-understood\cite{lykken}.
Experience with similar jets$+\eslt$ searches at the Fermilab Tevatron suggest that it may well
take some time to fully understand detector performance, so that $\eslt$ can be reliably 
measured. For this reason, several of us recently proposed that {\it early} searches
for SUSY matter at the LHC may be better served by looking for events containing
multiple (2,3,4,...), high transverse momentum ($p_T\ge 20$ GeV)  isolated {\it leptons} 
($e$s and/or $\mu$s) along with jets, {\it instead} of $\eslt +$jets events\cite{bps}.
Requiring high lepton multiplicity rejects SM background at a large rate, while
maintaining much of the expected signal, since isolated leptons are expected to be produced 
frequently in the sparticle cascade decays\cite{multi}.

Since publication of Ref. \cite{bps}, it has been pointed out that reliable {\it electron}
identification may also be a major issue during the early phase of LHC running. If so, 
this could jeopardize the results of Ref. \cite{bps}, which summed over both muons and electrons
in order to establish the multi-lepton signal and background rates.
In addition, the SM background calculation of Ref. \cite{bps} included only 
$2\to 2$ processes that were pre-programmed into Isajet.
However, various SM $2\to n$ BG processes potentially may be larger than the lowest order
processes considered in Ref. \cite{bps}.

In this letter, we show {\it i}). that it is sufficient to focus only on isolated {\it multimuon} plus jets events
during the earliest SUSY searches at LHC. The lack of electron channels can be partially compensated
for by the lower $p_T$ values which are allowed for isolated muon searches. Secondly, {\it ii}). we evaluate
a variety of additional $2\to n$ background processes beyond those presented in Ref. \cite{bps},
thus putting our results on a more firm foundation.
Thirdly, we re-evaluate all signal and background channels for the anticipated start-up
energy of $\sqrt{s}=10$ TeV, instead of design energy $\sqrt{s}=14$ TeV.
Finally, {\it iii}). we scan over a wide swath of mSUGRA model parameter space, and present the LHC reach 
plot on the m(squark) vs. m(gluino) plane for various low levels of integrated luminosity.
In the same-sign dimuon plus jets channel, some reach is possible even for integrated luminosities as low as
0.1 fb$^{-1}$, where squark and gluino masses up to $\sim 450$ GeV may be probed.

There are several advantages to
a SUSY search via multimuon plus jets events. 
\bi
\item Reliable electron identification may be difficult in the early stages of
LHC running. Electrons will need to be readily distinguishable from QCD jets and also
from high $p_T$ photon production. As an example, a jet with a single soft charged pion  
plus several $\pi^0$s can give a track pointing to a mainly electromagnetic
calorimeter deposition, which may well fake an electron signal.
\item Muon identification should be straightforward even in the very early stages of LHC
running\cite{muonLHC}. In fact, cosmic ray muons have already been seen at both Atlas and CMS.
Muons with $p_T\agt 5$ GeV should readily penetrate the electromagnetic calorimeter (ECAL) 
and hadronic calorimeter (HCAL), yielding easily-seen tracks in the muon chambers.
Since muons are so heavy, they produce minimal bremsstrahlung and showering in the ECAL or HCAL.
\item Muons can be readily identified at $p_T$ values much lower than electrons.
Reliable $e$ tagging typically needs $p_T(e)\agt 20$ GeV, while $p_T(\mu )\agt 5$ GeV
is sufficient for muon identification. Thus, the lower $p_T$ muons emerging from
cascade decays will be easily detected, while this is not so for electrons.
\item Superparticle cascade decays tend to be rich in $b$s and $\tau$s. 
While $b\to c\mu\bar{\nu}_\mu$ decay yields mainly non-isolated muons, 
$\tau\to\nu_\tau\mu\bar{\nu}_\mu$ decay leads to rather soft, but isolated, muon
production. The rather low $p_T(\mu )$ requirements allows one to detect muons from $\tau$ decay, 
while $e$s from tau decay are often too soft to reliably identify.
\ei

The large rate for $b$ and $\tau$ production in sparticle cascade decay events has three sources\cite{ltanb}:
1. the large $b$ and $\tau$ Yukawa couplings, especially at large $\tan\beta$ values,
enhance chargino and neutralino branching fractions into $b$ and $\tau$ states, 
2. third generation
sparticle masses are often much lighter than their first/second generation counterparts due
to Yukawa coupling effects pushing the third generation SSB masses to low values, and also due
to large mixing effects, which are proportional to the corresponding {\it fermion} mass. This latter
effect also enhances sparticle decay rates into third generation fermions. 3. Higgs bosons, especially
$h$, can be produced at large rates in sparticle cascade decays. For instance, if the decay 
$\tz_2\to\tz_1 h$ is kinematically allowed, this usually dominates the $\tz_2$ branching fraction.
Since $h$ and the other Higgs subsequently decay dominantly into third generation fermions,
one gets enhanced $b$ and $\tau$ production from cascade decays of sparticles into Higgs bosons. 

The search for multi-muon events has been proposed much earlier with regards
to the search for fourth generation quarks\cite{bbgp}, which also decay via a cascade to the lightest
flavor states. Multimuon detection has been proposed in the old idea of an ``iron ball detector'',
wherein the interaction region is completely surrounded by iron absorber, and one only detects the 
penetrating muons\cite{ib}. LHC detectors are vastly more complex than the iron ball detector. But in the very
early stages of running, wherein calorimeter and other detector response is not well understood, 
their initial performance may approximate the iron-ball idea.

We adopt the Isajet 7.78 program for sparticle mass calculations and 
simulation of signal events at the LHC\cite{isajet}. 
A toy detector simulation is employed with
calorimeter cell size
$\Delta\eta\times\Delta\phi=0.05\times 0.05$ and $-5<\eta<5$. 
(here, $\eta =-\log\tan{\theta\over 2}$ is pseudorapidity and $\phi$ is
angle transverse to beamline). The HCAL
(hadronic calorimetry)
energy resolution is taken to be $80\%/\sqrt{E}+3\%$ for $|\eta|<2.6$ and
FCAL (forward calorimetry) is $100\%/\sqrt{E}+5\%$ for $|\eta|>2.6$. 
The ECAL (electromagnetic calorimetry) energy resolution
is assumed to be $3\%/\sqrt{E}+0.5\%$. We use the Isajet\cite{isajet} 
jet finding algorithm
with jet cone size $R\equiv\sqrt{\Delta\eta^2+\Delta\phi^2}=0.4$ 
and require that $E_T(jet)>50$ GeV and
$|\eta (jet)|<3.0$. Muons are considered
isolated if they have $p_T(\mu )>5 $ GeV and $|\eta_\mu|<2$ with 
visible activity within a cone of 
$\Delta R<0.2$ of
$\Sigma E_T^{cells}<5$ GeV. The isolation criterion helps reduce
multi-lepton backgrounds from heavy quark ($c\bar c$ and $b\bar{b}$) production.

%We identify a hadronic cluster with $E_T>50$ GeV and $|\eta(j)|<1.5$
%as a $b$-jet if it contains a $B$ hadron with $p_T(B)>15$ GeV and
%$|\eta (B)|<3$ within a cone of $\Delta R<0.5$ about the jet axis. We
%adopt a $b$-jet tagging efficiency of 60\%, and assume that
%light quark and gluon jets can be mis-tagged as $b$-jets with a
%probability $1/150$ for $E_T\le 100$ GeV, $1/50$ for $E_T\ge 250$ GeV, 
%with a linear interpolation for $100$ GeV$<E_T<$ 250 GeV\cite{xt}. 

For our initial analysis, we adopt the well-studied \sps1ap benchmark
point\cite{snops}, which occurs in the minimal supergravity (mSUGRA) model
with parameters $m_0=70$ GeV, $m_{1/2}=250$ GeV, $A_0=-300$ GeV, 
$\tan\beta =10$, $\mu >0$ and $m_t=172.6$ GeV.
The \sps1ap point leads to a spectrum with $m_{\tg} =608$ GeV, while squark masses
tend to be in the 550 GeV range. The gluinos and squarks then cascade decay via
a multitude of modes leading to events with high jet, $b$-jet, isolated lepton
and tau lepton multiplicity.

\begin{figure}[tb]
\includegraphics[width=.68\textwidth]{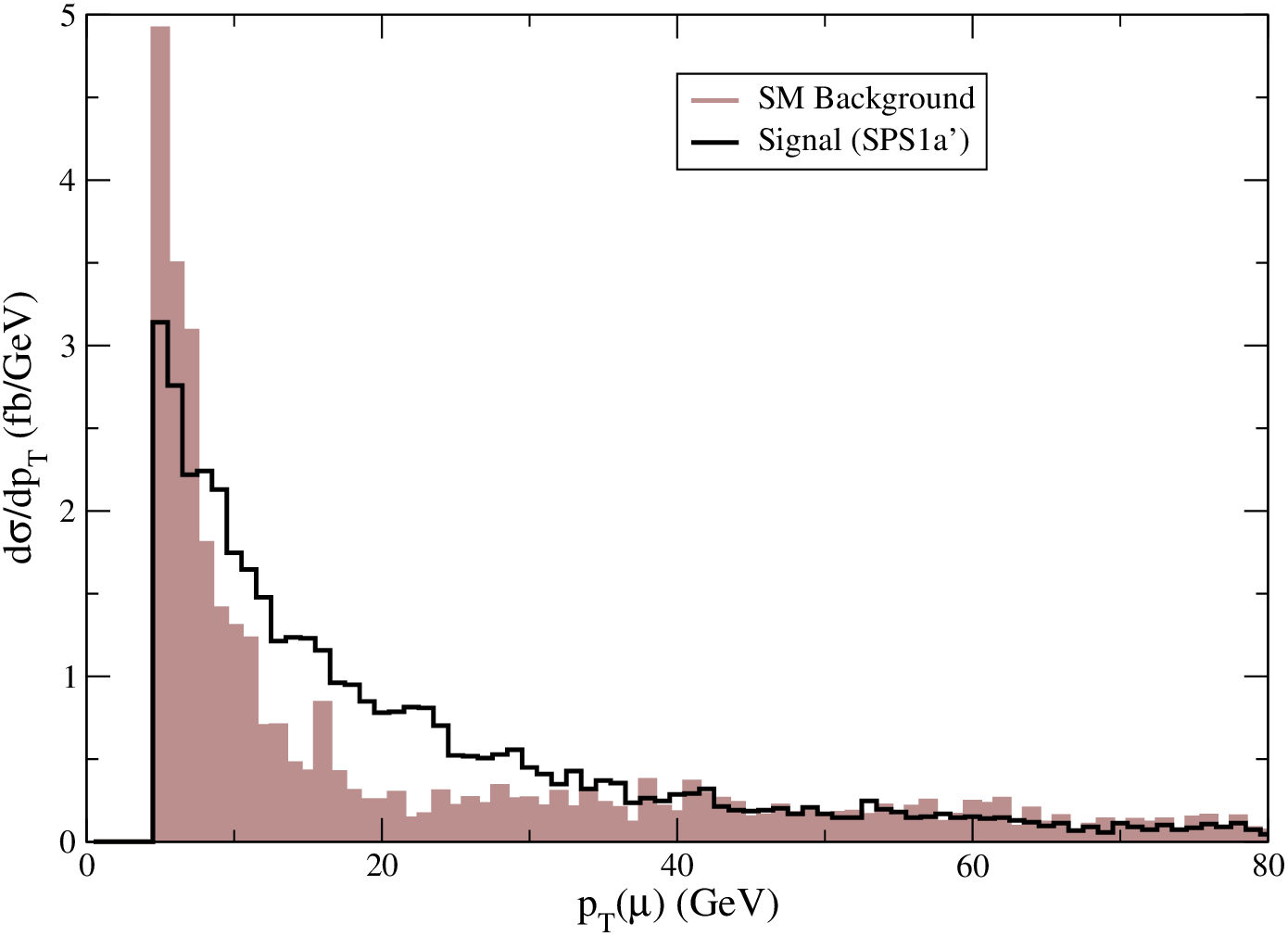}
\caption{\label{fig:ptmu} {\it $p_T(\mu )$ distribution
from the \sps1ap mSUGRA study, along with SM BG,
for SS dimuon plus $\ge 4$ jet events at LHC with $\sqrt{s}=10$ TeV. 
}}
\end{figure}

Since the gluino and squark cascade decay events will be rich in
jet activity, we first require events with 
$\ge 4$ jets, with $E_T(j1,j2,j3,j4) \ge  100,\ 50,50,50\ {\rm GeV}$.
We also require sphericity (restricted to the transverse plane) 
$S_T\ge 0.2$ to reject QCD-like
events at little cost to signal.
We {\it do not} apply the traditional cut on missing transverse energy, 
since at this stage we are working towards early SUSY discovery, 
when $\eslt$ may not yet be well-established.
In Fig. \ref{fig:ptmu} we show the muon $p_T$ distribution from point \sps1ap along with 
dominant SM BGs (see discussion below)
in same-sign (SS) dimuon plus $\ge 4$ jet events (each muon in a SS-dimuon event 
will have an entry in the plot). The signal muons tend to populate the $5-40$ GeV 
regime, and so should be easily measured by the bend of their tracks in the detector 
magnetic field. The BG muons come dominantly from $t\bar{t}$ production, and have a hard 
component (from $W$ decay) and a soft component (from rare $b$ and $c$ decays to isolated muons). 
The soft component exceeds signal in the 5-10 GeV range. 
Hence, we require $p_T(\mu )>10$ GeV in our multi-muon signal events, which eliminates 
much of the BG from $t\bar{t}$ production.

\begin{figure}[tb]
\includegraphics[width=.68\textwidth]{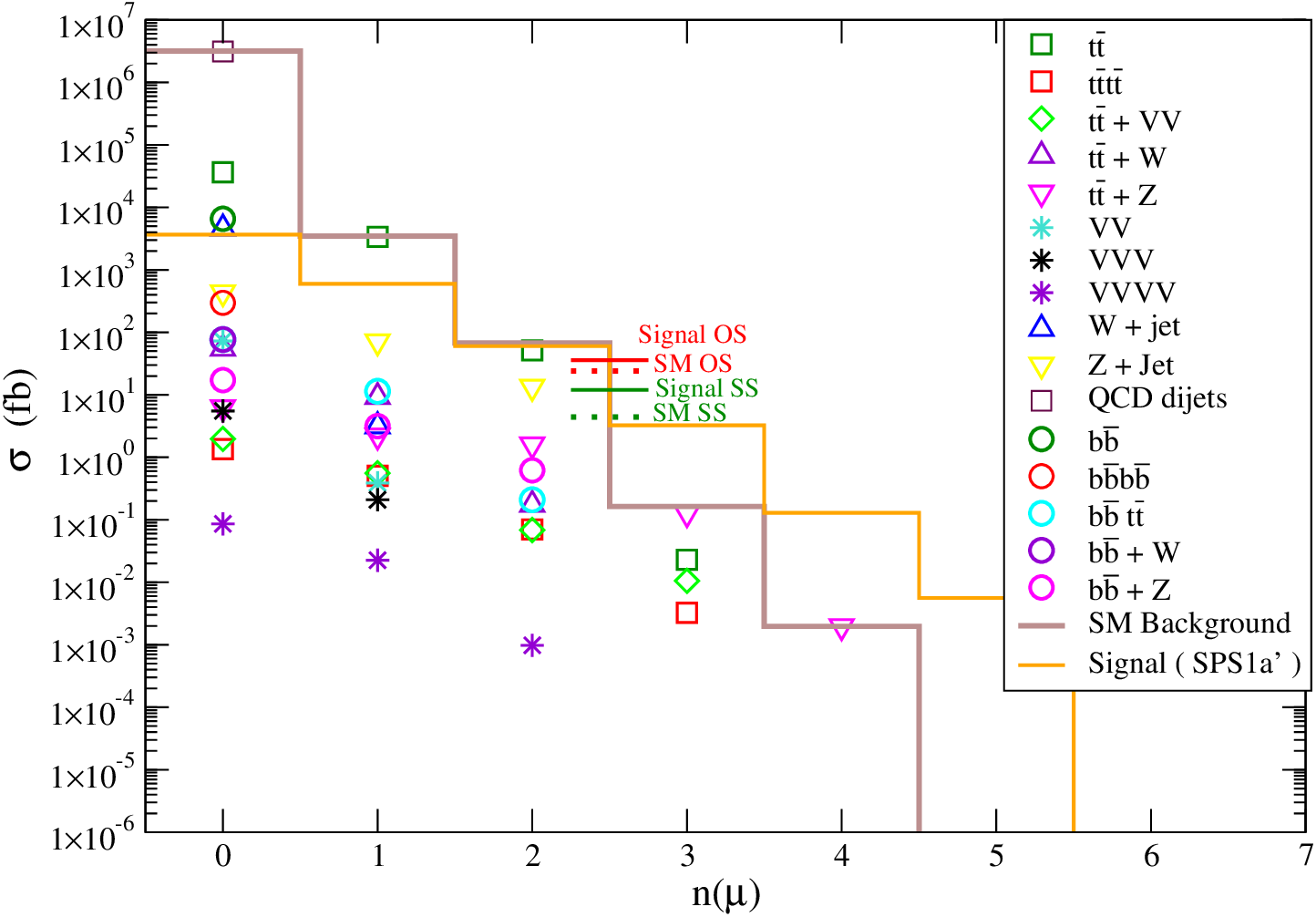}
\caption{\label{fig:nmu} {\it Muon multiplicity cross sections 
expected from the \sps1ap mSUGRA case study, along with SM background,
for $\ge 4$ jet plus $n$-muon events ($p_T(\mu )\ge 10$ GeV) at LHC with
$\sqrt{s}=10$ TeV. 
}}
\end{figure}

Next, we plot the multiplicity of muons in the SUSY cascade decay events.
The results are shown for point \sps1ap in Fig. \ref{fig:nmu} for $pp$ collisions
at $\sqrt{s}=10$ TeV.
We also plot a variety of SM backgrounds. The dominant backgrounds for the dimuon signal 
were calculated using AlpGen/Pythia and AlpGen's matching algorithm (MLM scheme\cite{alpgen}) 
to include multiple jet emission.
In particular, the $t\bar{t}$ channel includes $t\bar{t}+0,1,2,3$ jets, 
$Z+jet$ includes $Z+0,1,2,3$ jets, the $t\bar{t}+Z$ channel includes $t\bar{t}+Z+0,1,2$ jets
and $b\bar{b}+Z$ includes $b\bar{b}+Z+0,1,2$ jets (in all cases the full matrix element
$\gamma^*,Z^*\rightarrow l^+l^-$ was used). The presence of hard additional jets increases the
BG quite a bit from our earlier estimates using just the Isajet parton shower.\footnote{
Our BG from $b\bar{b}$ production comes from Alpgen/Pythia, but with just LO $b\bar{b}$
production along with jets from the parton shower.
We expect the BG from $b\bar{b}+$jets production to be sub-dominant because we would need to obtain one isolated
lepton from a $b$ decay, and another from a $c$ decay, while producing four hard jets at the same time,
and that sort of event is extremely rare. This reaction, with exact 4-jet emission matrix elements,
is extremely hard to generate with reliable statistics.}

In addition, we have calculated using MadGraph/Pythia \cite{madgraph,pythia} a variety
of exact $2\to n$ processes: $t\bar{t}t\bar{t}$, $t\bar{t}b\bar{b}$, $b\bar{b}b\bar{b}$, 
$t\bar{t}V$, $t\bar{t}VV$, $VV$, $W+jet$, $b\bar{b}$, QCD dijets,
$VVV$ and $VVVV$ production, where $V=W^\pm$ or $Z^0$. The summed BG histogram along with 
component contributions are also shown in Fig. \ref{fig:nmu}. 
We adopt a renormalization/factorization scale choice $Q=\sqrt{\hat s}/6$ 
($\hat{s}$ is the parton-parton CM frame squared energy) which brings our 
background (BG) cross
sections into close accord with NLO QCD results.\footnote{The dominant BG for SS dimuon plus
$\ge$ 4-jet events comes from $t\bar{t}$ production. Using the MCFM code\cite{mcfm},
we find $\sigma^{LO} (pp\to t\bar{t} )(Q=m_{top})\simeq 255$ pb, while 
$\sigma^{NLO} (pp\to t\bar{t} )(Q=m_{top})\simeq 347$ pb. If instead we take $Q=\sqrt{\hat{s}}/6$, 
then $\sigma^{LO} (pp\to t\bar{t}X )(Q=\sqrt{\hat{s}}/6)\simeq 337$ pb.}

At $n(\mu )=0$, 
signal is about three orders of magnitude below SM background. As we increase
the isolated muon multiplicity, BG falls off faster than signal, and signal exceeds
BG already at $n(\mu )=3$, where signal is at the $\sim 5$ fb level.
The dimuon signal can be broken up into opposite sign $\mu^+\mu^-$ events (OS)
and same sign $\mu^\pm\mu^\pm$ events (SS)\cite{ssleps,multi,mitsel}. In order to suppress
the large contribution from the $Z$ peak to the OS events we required 10 GeV$\le m(\mu^+\mu^-) \le 75$ GeV.
The SS signal is due in part to the Majorana nature of the gluinos, in that 
a gluino is as likely to decay via $\tg\to\tw_1^+ \bar{u}d $ as via 
the charge conjugate mode. Thus, $\tg\tg$ production is likely to lead to equal amounts of
$++$ and $--$ SS dileptons. Now, $\tg\tq$ or $\tq\tq$ production depends on the quark content of the colliding
beams, and since LHC is a $pp$ collider, we expect more $++$ dileptons than $-\ -$ dileptons from
squark production.

For OS dimuons and case study \sps1ap ,
the OS signal is just slightly above OS BG, while SS signal well exceeds SS BG, and is at
the $\sim 10$ fb level. As we move to higher and higher muon multiplicity, 
the signal rates diminish, although signal-to-background ratio steadily improves.
For instance, at $n(\mu )=3$, signal is $\sim 5$ fb, while the summed SM background, 
arising mainly from $t\bar{t}$ and $t\bar{t}Z$ production, occurs at the
$\sim 0.1$ fb level.
Using these results, we can now see that if case study \sps1ap describes SUSY, then the first
clear signal may emerge in the SS dimuon plus multi-jet channel, with corroborating signals in 
the OS and tri-muon channels.

\begin{figure}[tb]
\includegraphics[width=.68\textwidth]{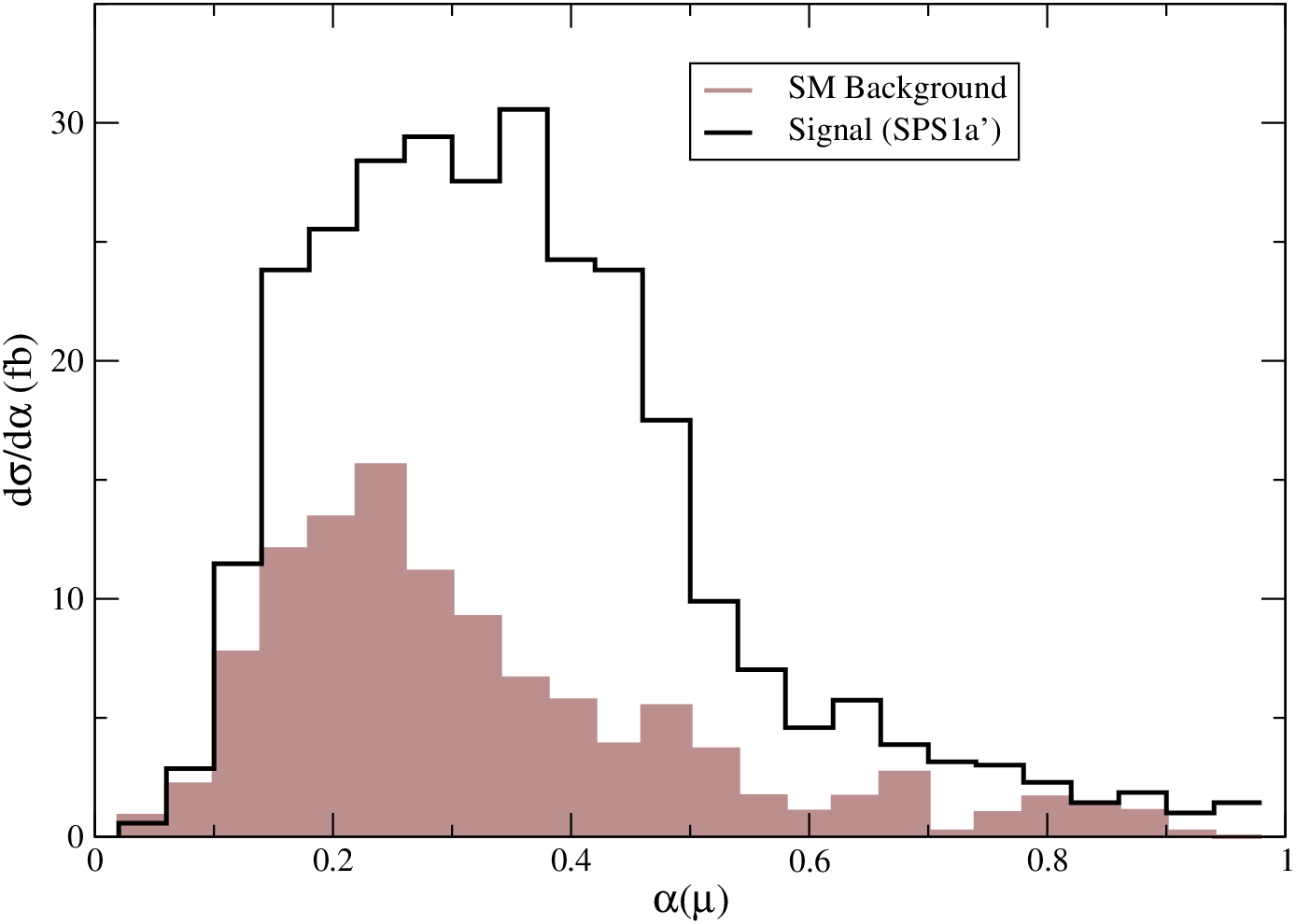}
\caption{\label{fig:alpha} {\it Plot of $\alpha =p_T(\mu_2 )/m(\mu^\pm\mu^\pm )$
for SS dimuon plus jets events at LHC with $\sqrt{s}=10$ TeV, after cuts listed in text.
}}
\end{figure}

Another discriminating variable for SUSY events has been proposed 
by Randall and Tucker-Smith\cite{rt}, 
albeit applied to dijet events coming from squark pair production. They propose
using $\alpha=p_T(jet_2)/m(jet_1,jet_2)$. Here we plot $\alpha(\mu)=p_T(\mu_2)/m(\mu^\pm\mu^\pm)$ 
in Fig. \ref{fig:alpha},
and do find that the SUSY event shape is discriminated from the SM event shape.
We do not apply an $\alpha(\mu)$ cut at this time for very low luminosity studies, but merely note
that this distribution will add additional confidence in any possible 
SS dimuon signal. 
%

%The OS dimuon channel should also be visible, 
%and will be characterized by a well-known kinematic edge at 
%$m(\mu^+\mu^- )\le m_{\tz_2}-m_{\tz_1}\simeq 85$ GeV, which occurs when the second lightest neutralino 
%$\tz_2$ decays to $\mu^+\mu^- \tz_1$ or $\tmu^\pm\mu^\mp$ states\cite{mlledge}. 
%Simulated background and signal histograms are shown in Fig. \ref{fig:mll}, where the edge
%just below the $Z$ peak (at $M_Z\simeq 91$ GeV) is visible.
%
%\begin{figure}[tb]
%\includegraphics[width=.48\textwidth]{invmass.eps}
%\caption{\label{fig:mll} {\it OS dimuon invariant mass for
%the \sps1ap mSUGRA case study along with SM background,
%for $OS$ dimuon plus $\ge 4$-jet events at LHC. 
%}}
%\end{figure}
%

As higher integrated luminosities are reached, trimuon and later four muon plus jet events should emerge
at rates far above expected background. Of course, also as higher integrated luminosities are
achieved, reliable electron ID should become available, and ultimately also reliable $\eslt$ measurements.
Thus, the real utility of multi-muon plus jets events will be for a possible {\it early} 
discovery of SUSY, when muon ID is possible, but electron ID and $\eslt$ resolution are still
works in progress.

\begin{figure}[tb]
\includegraphics[width=.68\textwidth]{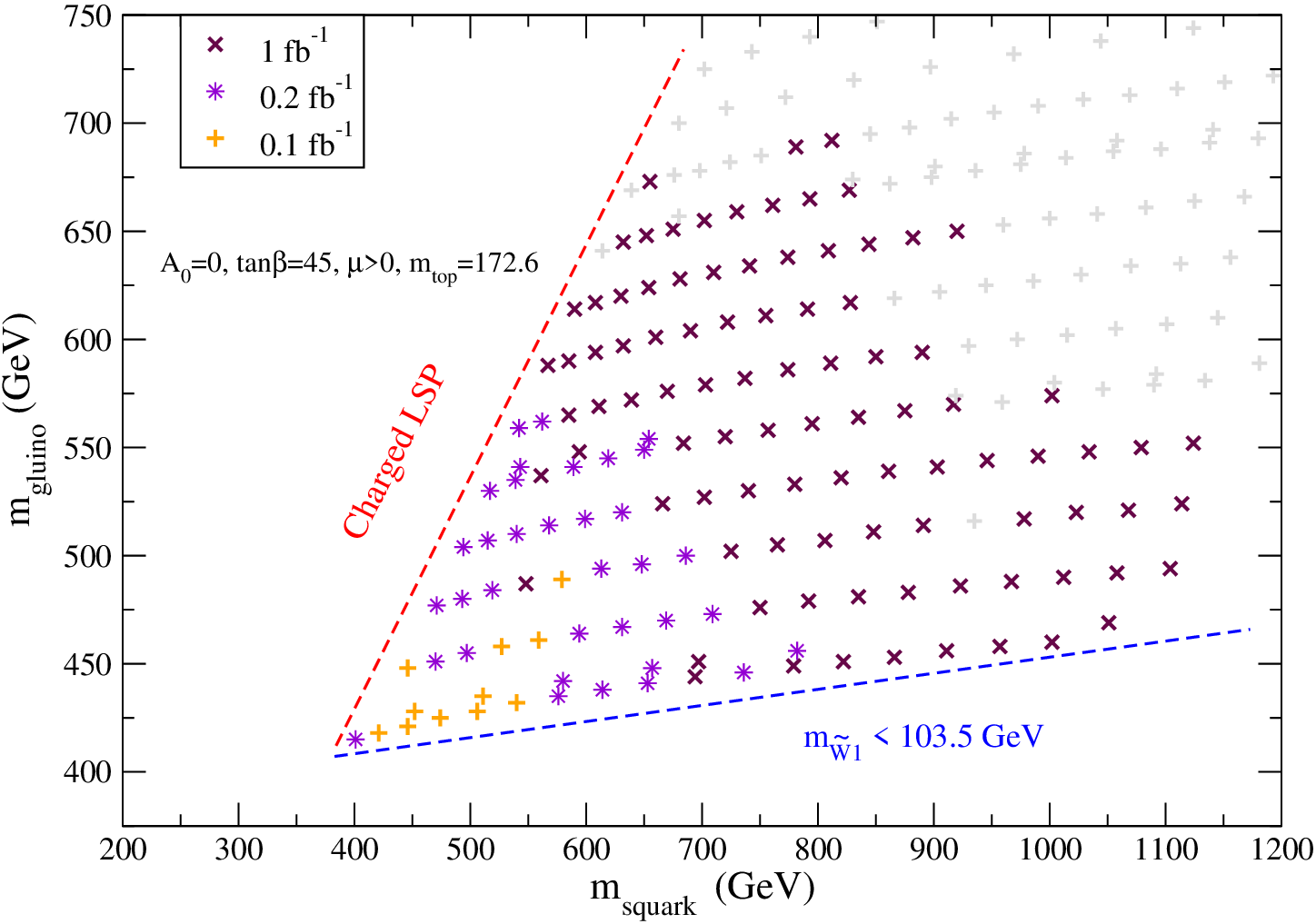}
\caption{\label{fig:reach45} {\it Reach of the $\sqrt{s}=10$ TeV LHC for mSUGRA
models with $A_0=0$ and $\tan\beta =45$ via SS dimuon $+\ge 4$ jet events 
in the $m_{\tg}$ vs. $m_{\tq}$ plane, for
various integrated luminosity values.
}}
\end{figure}

In Fig. \ref{fig:reach45}, we scan over $\sim 200$ choices of $m_0$ and $m_{1/2}$ values 
for fixed $A_0=0$, $\mu >0$ and $\tan\beta =45$. We test to see if the SS dimuon
plus jets signal is greater than a nominal discovery threshold of $5\sigma$, and require
at least five signal events as well, for various integrated luminosity choices:
$0.1,\ 0.2$ and 1 fb$^{-1}$. We plot the results in the physical
$m_{\tg}$ vs. $m_{\tu_L}$ plane. The lower-right region gives a chargino mass
less than 103.5 GeV, and so is already excluded by LEP2 new particle searches. The
left side of the plot gives a $\ttau_1$ slepton as the LSP, and is excluded by
null searches for stable, charged relics from the Big Bang. 
For just  0.1 fb$^{-1}$ of integrated luminosity, eleven
points are accessible, with $m_{\tg}\alt 480$ GeV and $m_{\tu_L}\alt 580$ GeV.
For 0.2 fb$^{-1}$, $m_{\tg}\alt 550$ and $m_{\tq}\alt 700$ GeV are being probed.
The SS di-muon reach increases to $m_{\tg}\sim 650$ GeV for 1 fb$^{-1}$ of integrated
luminosity.
If we move to a lower $\tan\beta =10$ value,
then the dimuon reach diminishes only slightly from that presented in the
$\tan\beta =45$ case.
%multi-muon production from stau decays increases the reach somewhat, and even larger
%values of $m_{\tg}$ and $m_{\tu_L}$ can be probed.
%
%\begin{figure}[tb]
%\includegraphics[width=.48\textwidth]{Sq_gluinoC_ss.eps}
%5\caption{\label{fig:reach10} {\it Reach of the LHC for mSUGRA
%models with $A_0=0$ and $\tan\beta =10$ via SS dimuon $+\ge 4$ jet events 
%in the $m_{\tg}$ vs. $m_{\tq}$ plane, for
%various integrated luminosity values.
%}}
%\end{figure}
%
%

{\it Conclusions:} In the early stages of LHC running, electron ID, ECAL and HCAL
calibration and $\eslt$ resolution may all be {\it works in progress}. However,
muon ID and momentum resolution, obtained from track bending in the magnetic field,
should be quite reliable, and allow muon $p_T$ measurement down to the $\sim 5-10$ GeV
range. SS dimuon and, later, OS dimuon and $\ge 3\mu$ plus multi-jet signals, without any $\eslt$ discrimination,
should allow for good signal-to-background resolution for gluino masses 
up to about 550 GeV with just 0.2 fb$^{-1}$ of data. Thus, SS dimuon and multi-muon 
plus jets production
offer excellent possibilities for an early SUSY discovery at LHC, even if $\eslt$ and electron ID
are not initially well-understood.

\begin{acknowledgments}
We thank S. Rydbeck, M. Strauss and G. Mitselmaker for discussions.
\end{acknowledgments}


\end{document}